\documentclass[a4paper]{aa}
\usepackage{graphicx}
\usepackage{txfonts}
\usepackage{natbib}
\bibpunct{(}{)}{;}{a}{}{,} 
\usepackage{bm}

\begin{document}
\title{The kinematics of NGC1333-IRAS2A -- a true Class 0 protostar}
\titlerunning{The kinematics of NGC1333-IRAS2A}

\author{
  C.~Brinch\inst{1,2} \and 
  J.~K.~J\o rgensen\inst{2} \and
  M.~R.~Hogerheijde\inst{1}}  
\date{}
\institute{
  Leiden Observatory, Leiden University, 
  P.O.~Box 9513, 2300 RA Leiden, The Netherlands\\ 
  \email{cbrinch@astro.uni-bonn.de} \and 
  Argelander-Institut f\"ur Astronomie, Universit\"at Bonn,
  Auf dem H\"ugel 71, 53121 Bonn, Germany}

\abstract
{Low-mass star formation is described by gravitational collapse of dense cores of gas and dust. At some point during the collapse, a disk is formed around the protostar and the disk will spin up and grow in size as the core contracts because of angular momentum conservation. The question is how early the disk formation process occurs.}
{In this paper we aim to characterize the kinematical state of a deeply embedded, Class 0 young stellar object, NGC1333--IRAS2A, based on high angular resolution ($< 1\arcsec \approx 200$ AU) interferometric observations of HCN and H$^{13}$CN $J=$ 4--3 from the Submillimeter Array, and test whether a circumstellar disk can be detected based on gas kinematic features.}
{We adopt a physical model which has been shown to describe the object well and obtain a fit of a parameterized model of the velocity field, using a two-dimensional axis-symmetric radiation transfer code. The parameterization and fit to the high angular resolution data characterize the central dynamical mass and the ratio of infall velocity to rotation velocity.}
{We find a large amount of infall and very little rotation on all scales. The central object has a relatively low mass of 0.25 M$_\odot$. Our best fit is consistent with both the interferometer data and single-dish observations of the same molecule.}
{As an object with a low stellar mass compared to the envelope mass, we conclude that NGC1333--IRAS2A is consistent with the suggestion that, as a Class 0 object, it represents the earliest stages of star formation. The large amount of infall relative to rotation also suggests that this is a young object. We do however find the need of a central compact component on scales of a few hundred AU based on the continuum data, which suggests that disk formation happens shortly after the initial gravitational collapse. The data do not reveal a distinct velocity field for this 0.1 M$_\odot$ component. }
\maketitle

\section{Introduction}
An open question in the subject of low-mass star formation is how and when the velocity field of the molecular gas in the protostellar envelope changes from being dominated by radial infall to being dominated by Keplerian rotation. In the early stages of star formation, gas falls toward the center as gravitational collapse sets in~\citep{shu1977}. Cores, however, are known to carry a certain amount of rotational energy~\citep{goodman1993}, so as a core contracts, the net angular momentum spins it up, flattening the infalling material in the direction of the rotation axis~\citep{cassen1981,terebey1984, adams1988,basu1998}. After the main accretion phase, the envelope is gone and a planet forming disk with gas in Keplerian orbits is left. 

In this paper, we study the kinematics of the deeply embedded NGC1333-IRAS2A protostar using subarcsecond resolution data from the Submillimeter Array (SMA) as well as detailed two-dimensional line radiative transfer and the model framework presented by \citet{brinch2008a}. In that paper it was shown that the evolution of the velocity field of a hydrodynamical simulation of star formation, from the collapse phase to the disk phase, can be approximated by a single parameter, namely the radially averaged ratio of infall to rotation. Under the assumption that the simulation used in that work provides a good approximate description of the formation of stars, this parameter may be linked to the age of a protostar and can thus tell us about the evolutionary stage of the system.

Observationally, young stellar objects are classified using the scheme laid out by \cite{lada1984} with the addition of the Class 0 by \cite{andre1993}. In this scheme, the evolution of the young star is related to the shape of the spectral energy distribution with the addition that Class 0 objects are defined as protostars with $M_*<$ $M_{\rm{env}}$, which is equivalent to the original empirical definition of Class 0 objects as stars with a high submillimeter flux relative to their bolometric luminosity. The mass of the central object is not easily measured, but one way to estimate it is by determining the velocity field, since the magnitude of the velocity vectors are determined by the central mass {\bf when the material becomes dominated by rotation}.

NGC1333-IRAS2 was first detected through IRAS observations by \cite{jennings1987}. Submillimeter continuum imaging has subsequently shown that it likely consists of three different objects, NGC1333-IRAS2A, -IRAS2B, and -IRAS2C, of which the two former are well isolated and detected at mid-IR wavelengths, e.g., with the Spitzer Space Telescope \citep{jorgensen2007a,gutermuth2008}. These sources also show compact emission in millimeter interferometric images \citep{looney2000,jorgensen2004}. The third source, IRAS2C, shows relatively strong N$_2$H$^+$ emission possibly reflecting the absence of a central heating source \citep{jorgensen2004}. Modeling of the high angular resolution millimeter continuum observations \citep{looney2003,jorgensen2004} revealed the existence of a central compact continuum component in NGC1333-IRAS2A, likely reflecting the presence of a central circumstellar disk with a mass of $\sim 0.1$~$M_\odot$ (depending on the assumed dust temperature and opacity). NGC1333-IRAS2A was targeted as part of the PROSAC survey of Class 0 embedded protostars \citep{jorgensen2005,jorgensen2007}. The 1\arcsec\ resolution submillimeter continuum data \citep{jorgensen2005} showed that the central compact component was resolved with a diameter of approximately 300~AU.


In this paper we investigate the kinematical properties of IRAS2A using a global parameterization of the velocity field. We are particularly interested in the velocity structure on scales $<$100~AU, traced by the high angular resolution Submillimeter Array\footnote{The Submillimeter Array is a joint project between the Smithsonian Astrophysical Observatory and the Academia Sinica Institute of Astronomy and Astrophysics and is funded by the Smithsonian Institution and the Academia Sinica.} data, where a circumstellar disk is thought to be present. The SMA instrument is ideal for this task because we can probe the warm, dense gas with high excitation lines at high angular resolution and thereby zoom in on the central parts of the object. We furthermore benefit from the fact that the interferometer is less sensitive to ambient cloud material.

This paper is laid out as follows: In Sects.~\ref{iras2:obs} and \ref{iras2:results} we present the data that we use, and our model and best fit are described in Sect.~\ref{iras2:analysis}. Discussion and a summary are given in Sects.~\ref{iras2:disc} and \ref{iras2:outlook}, respectively.

\section{Observations}\label{iras2:obs}
Supplementing the SMA compact configuration data from 2004 October 17 \citep{jorgensen2005}, NGC1333-IRAS2A was observed in the extended configuration on 2006 January 8 with seven of the eight antennas in the array at the time. The weather was good and stable during the observations with a sky opacity at 225 GHz of about 0.05. Combined with the 2004 compact configuration data, the observations provide good ($u,v$)-coverage on baselines between 17 and 210 k$\lambda$ with a few baselines at 232 to 264 k$\lambda$. The complex gains were calibrated by observations of the strong quasars 3c84 (2.2 Jy) and 3c111 (2.8 Jy) every 15 minutes. 

The SMA correlator was configured with an identical setup to the compact configuration observations with 2 GHz bandwidth at 345~GHz (LSB) and 355~GHz (USB). For the lines predominantly discussed in this paper, HCN and H$^{13}$CN $J$= 4--3 at respectively 354.5055 and 343.3398~GHz, the spectral resolution was 0.15~km~s$^{-1}$ (HCN) and 0.30~km~s$^{-1}$ (H$^{13}$CN). In the maps used in this analysis, we use the natural weighting scheme to optimize for signal-to-noise in the resulting image at a small cost of resolution. With natural weighting the RMS is 0.225 (HCN) and 0.147 (H$^{13}$CN) Jy beam$^{-1}$ in a 1 km~s$^{-1}$ wide channel with a beam size of 1.2\arcsec\ $\times$ 0.8\arcsec . The initial calibration of the data was done with the SMA MIR package \citep{qi2005} with additional image processing done using MIRIAD \citep{sault1995}.

In addition to these data, we use previously published HCN and H$^{13}$CN single-dish data from the James Clerk Maxwell Telescope \citep{jorgensen2004} as well as 3 mm interferometric observations from the BIMA array \citep{jorgensen2004b}.

\section{Results}\label{iras2:results}
\begin{figure}
  \begin{center}
    \includegraphics[width=8.5cm]{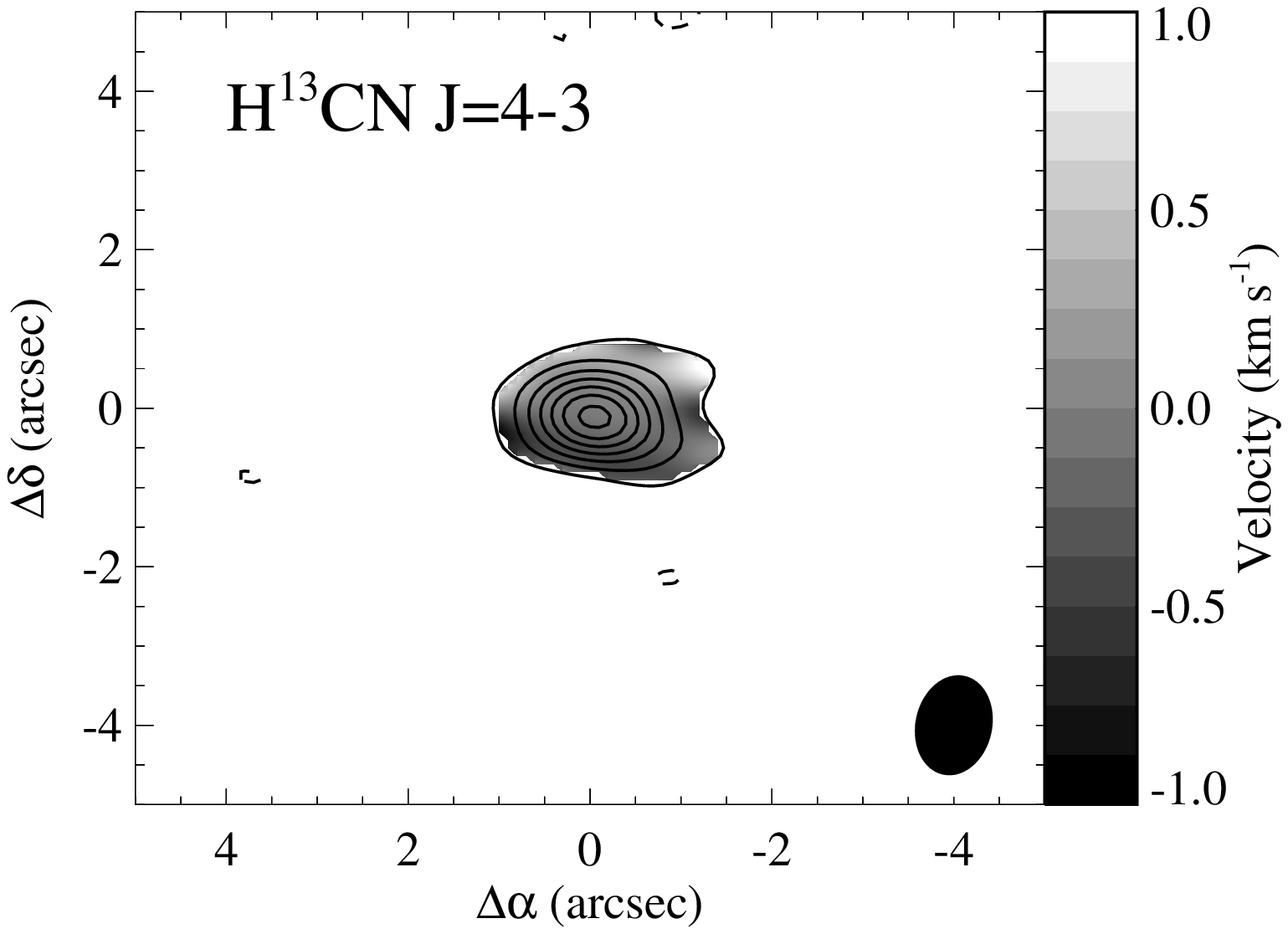}
    \includegraphics[width=8.5cm]{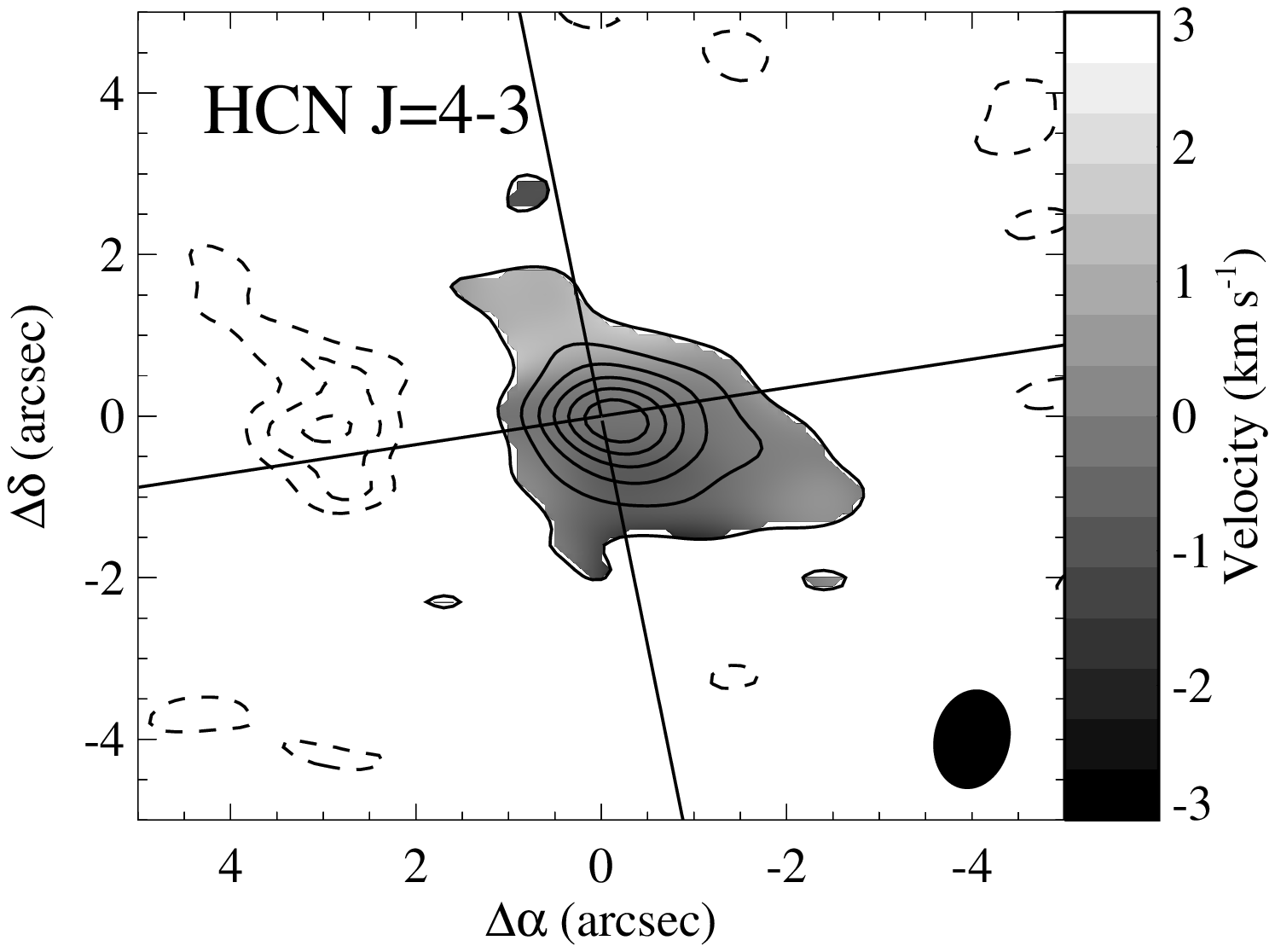}
  \end{center}
  \caption{Zero (contours) and first (grey scale) moments of H$^{13}$CN and HCN $J=$ 4--3. Contour lines showing the integrated intensity begin at 3$\sigma$ and increase by 3$\sigma$=2.3 Jy km~s$^{-1}$ (H$^{13}$CN) and 5.4 Jy km~s$^{-1}$ (HCN), respectively. The full lines indicate the directions of the outflows identified by \cite{knee2000}.}\label{moments}
\end{figure}

\begin{figure}
  \begin{center}
    \includegraphics[width=8.5cm]{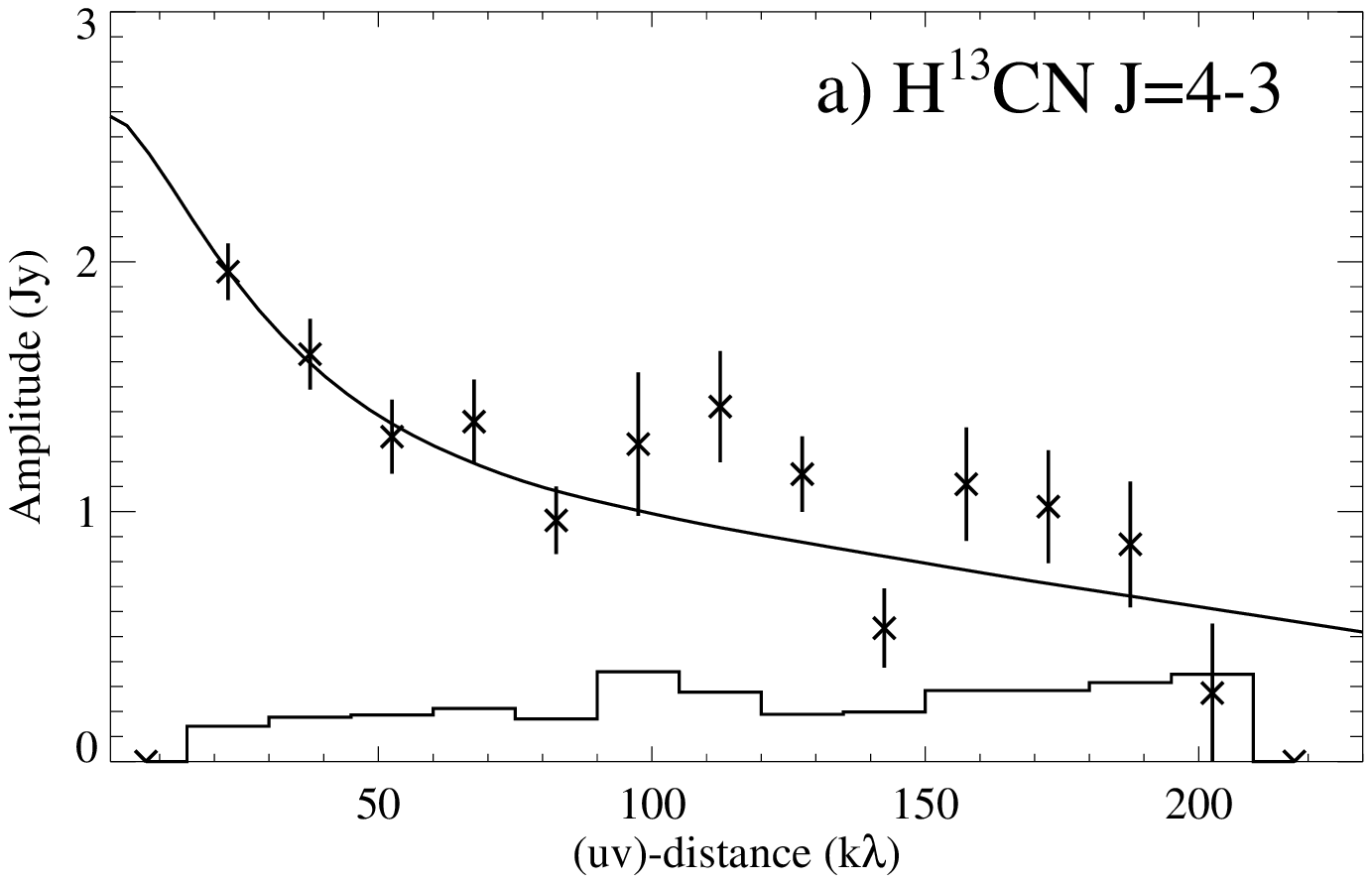}
    \includegraphics[width=8.5cm]{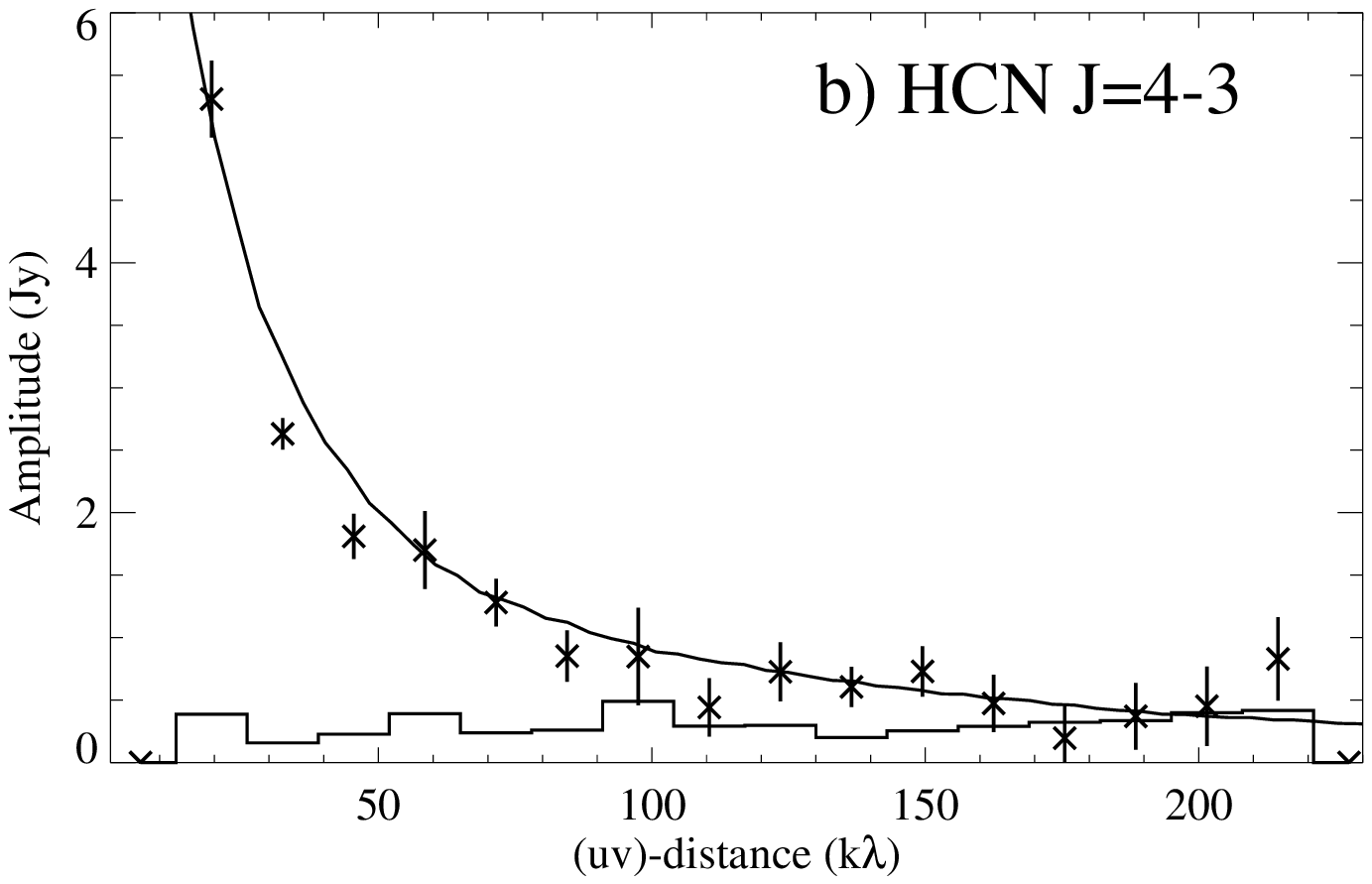}
  \end{center}
  \caption{Averaged ($u,v$)-amplitudes over the velocity range of $\pm$3 km~s$^{-1}$ from the systemic velocity of H$^{13}$CN and HCN. The histogram is the expected amplitude if no signal was present. The solid line is our best fit model as discussed in Sect.~\ref{iras2:analysis}.}\label{iras2:uvamps}
\end{figure}

Figure~\ref{moments} shows the zeroth and first moments of the image cubes, integrated intensity and velocity, respectively for both the H$^{13}$CN line and HCN line. In this figure, the contour lines show the zeroth moment and the grey scale shows the first moment. The contours start at the three sigma level corresponding to 5.4 Jy beam$^{-1}$ for HCN and 2.3 Jy beam$^{-1}$ for H$^{13}$CN and increase linearly by three sigma in both panels. The emission is seen to be centrally peaked with only a bit of extended emission at the 3--6 $\sigma$ level in HCN. This is in good agreement with the ($u,v$)-amplitude plots in Fig.~\ref{iras2:uvamps} where HCN is seen to be resolved while H$^{13}$CN is more compact. A small gradient in the velocity field {\bf is seen in the north-south direction in the HCN moment map with a magnitude of a few km~s$^{-1}$ over the 1000~AU scales of the observed emission}. A similar gradient {\bf was reported in emission on larger} scales by \cite{jorgensen2004b}. The gradient is most likely associated with {\bf one of the two outflows driven by IRAS2A (see panel b) Fig.~\ref{moments} --- corresponding to its direction and polarity. A similar gradient is not present in the H$^{13}$CN map. A fit of a linear gradient to this map shows that the systematic velocity difference across the source is at most 0.1 km~s$^{-1}$ over a scale of 2$''$ and that the direction of the gradient fit is highly dependent on the clip level of the map and thus unconstrained by the data. }

When the two SMA tracks are combined we get a good ($u,v$)-coverage on baselines between 17 and 210 k$\lambda$. This is equivalent to emission on scales of 0.9$''$ to 12.1$''$ which, at the distance of NGC1333 (220 pc; \citealt{cernis1990}), corresponds to linear scales of 198 to 2662 AU. Some flux is also measured on scales between 232 and 264 k$\lambda$ (167 to 190 AU), but at low signal-to-noise. Any emission on baselines shorter than 17 k$\lambda$ is filtered out by the interferometer. Figure \ref{iras2:uvamps} shows the averaged ($u,v$)-amplitudes as a function of baseline between 0 and 210 k$\lambda$. In panel a) is shown the H$^{13}$CN $J=$ 4--3 line. This panel shows that the emission on all scales up to 200 k$\lambda$ is largely compact with a strong increase in flux on scales below 50 k$\lambda$. The histogram indicates the expectation values if no signal were present and the full line is our best fit model, which we discuss in Sect.~\ref{iras2:analysis}. Panel b) has a similar layout, but for the HCN $J=$ 4--3 line. {\bf For this line the emission is also well resolved on baselines below 80 k$\lambda$, but with a lower signal-to-noise on longer baselines, disappearing in the noise at baselines longer than about 160~k$\lambda$. This is likely due to the HCN 4--3 emission becoming optically thick on small scales rather than HCN having an intrinsically different spatial distribution than H$^{13}$CN}.

\begin{figure}
  \begin{center}
    \includegraphics[width=8.5cm]{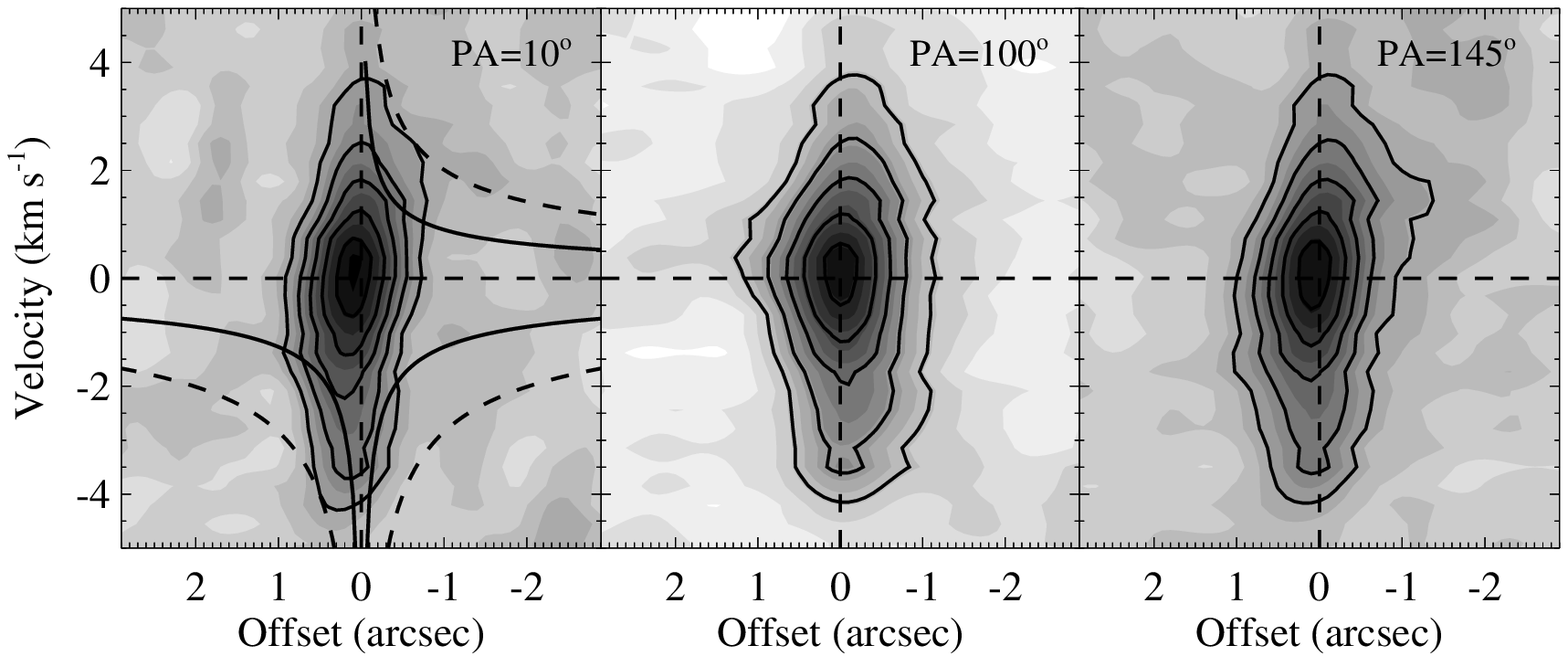}
    \includegraphics[width=8.5cm]{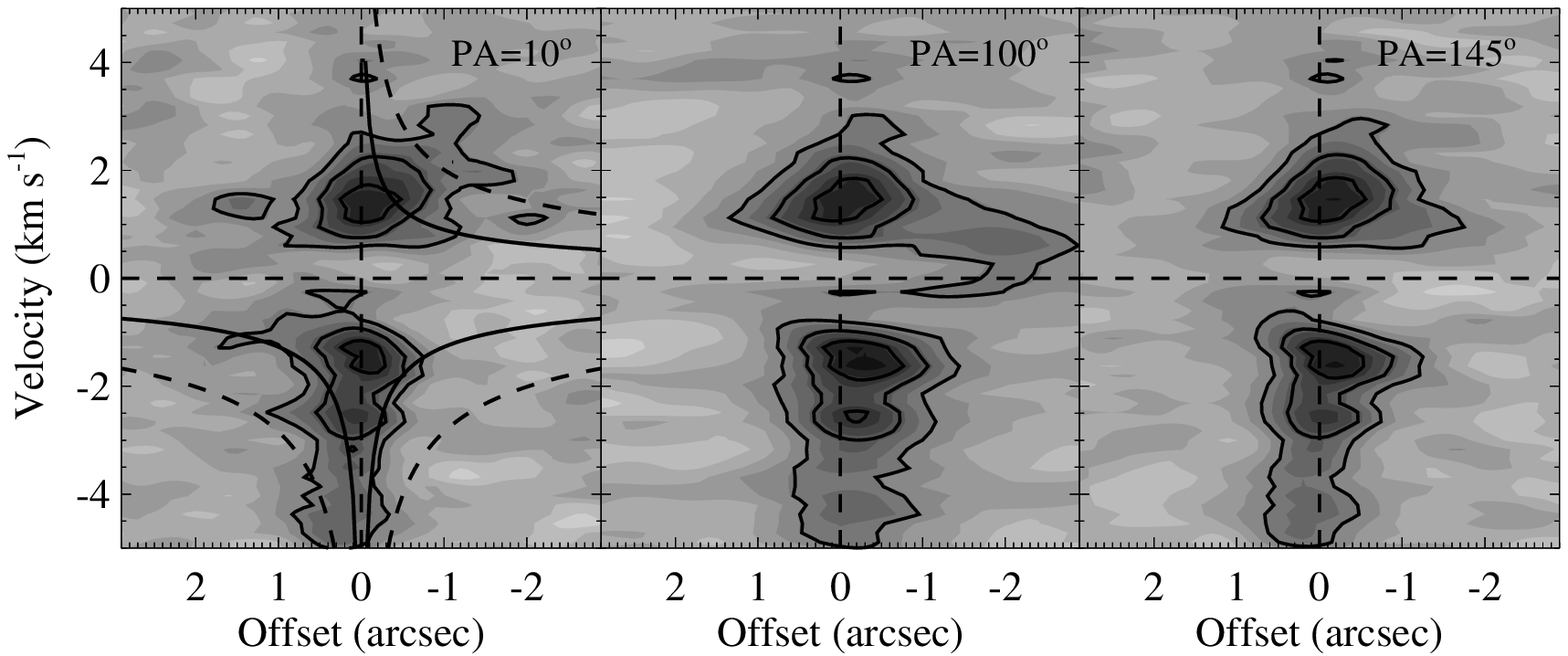}
  \end{center}
  \caption{PV-diagrams of the combined SMA data sets of H$^{13}$CN
    $J=$4--3 (top) and HCN $J=$4--3 (bottom). The three panels
    correspond to slices at different position angles with
    PA=10$^\circ$, PA=100$^\circ$, and PA=145$^\circ$. In the leftmost
    panels are shown Keplerian rotation (first quadrant) and free-fall
    (third and fourth quadrants) profiles for $M_*$=0.3 M$_\odot$
    (full line) and 1.0 M$_\odot$ (dashed line). The contour lines are
    shown as increments of 3$\sigma$ {\bf (0.9~Jy~arcsec$^{-2}$ and
      1.9~Jy~arcsec$^{-2}$ for H$^{13}$CN and HCN, respectively)} and have
    the same values in all three panels.}\label{pvs}
\end{figure}

Figure \ref{pvs} shows three position-velocity (PV) diagrams of both H$^{13}$CN and HCN. The PV-diagrams show the emission contours along one spatial axis and a range in velocities. The three panels of Fig.~\ref{pvs} correspond to PV-diagrams along slices of different position angles. If only rotation is present, the PV-diagram along the {\bf axis of rotation should show no evidence of systematic motions}, whereas the PV-diagram along the perpendicular axis should. The three axes for which we show PV-diagrams are chosen so that one coincides with the direction of the velocity gradient seen in Fig.~\ref{moments} (the same direction as the north-south outflow), one perpendicular to this (coinciding with the east-west outflow), and one which is inclined with 45$^\circ$ with respect to the other two axes (coinciding with the direction toward IRAS2B and IRAS2C). 

The first thing to notice is that the three {\bf position-velocity diagrams extracted for each of HCN and H$^{13}$CN along the different spatial axes appear quite similar}. There is no direction that obviously shows a clear rotation pattern and with {\bf the peak intensities at different velocities} located exactly at zero offset, there is actually very little sign of the velocity gradient at all, except for small skew in the contours in the left-most panel of H$^{13}$CN and emission at the 3--6 $\sigma$ level at positive velocities in the center HCN panel. The black lines that are also plotted in Fig.~\ref{pvs} show Keplerian rotation profiles for a central mass of 0.3 M$_\odot$ (full line) and 1.0 M$_\odot$ (dashed line) in the first quadrant. The curves in the two lower quadrants show free-fall velocity profiles for the corresponding central masses. Obviously, it is not possible to discriminate between rotation and infall by comparing the emission to these curves, because the emission is very compact, even for HCN. The emission contours are, however, more consistent with the 0.3 M$_\odot$ curves than with the 1.0 M$_\odot$ curves.  The second thing to notice in the PV-diagrams is how fast the emission drops off along the spatial axis while the emission at small offsets shows very broad lines. This is true for both HCN and H$^{13}$CN. It is therefore likely that the velocity gradient which is seen in the moment maps is due to the outflow rather than rotation. After all, this gradient does lie in the same direction as the outflow. These PV-diagrams also resemble the ones presented by \cite{hogerheijde1998} for a number of known outflow sources.

\section{Analysis}\label{iras2:analysis}
We describe the observed emission with the model for the density and temperature from~\citet{jorgensen2002}, who derived this model from analysis of the continuum image of the source at 450 and 850 $\mu$m as well as at its broadband SED from 60 $\mu$m to 1.3 mm. The density is described by a power-law,
\begin{eqnarray}
  n(r)=1.5\times 10^6 \textrm{\ cm}^{-3}(r/1000 \textrm{\ AU})^{-1.8},\label{iras2:density}
\end{eqnarray}
while the temperature was calculated using a 1D continuum radiation transfer code as described by \citet[see also Fig.~1f of J\o rgensen et al. 2004a]{jorgensen2002}. The outer radius of the model is $1.2\times 10^4$ AU. \citet{jorgensen2004} also modeled the chemical abundance profiles using the integrated intensity of the optically thin H$^{13}$CN single-dish lines. They found that to reproduce the line ratios, a model with a depletion zone is needed. The resulting HCN abundance profile reads,
\begin{eqnarray}
X(\textrm{HCN}) =  \Big \{
\begin{array}{l l}
2\times 10^{-8} & , n_{\textrm{H}_2} < 7\times 10^4 \textrm{\ cm}^{-3}\\
2\times 10^{-9} & , n_{\textrm{H}_2} > 7\times 10^4 \textrm{\ cm}^{-3}\\
\end{array},
\end{eqnarray}
assuming an isotopic abundance ratio of $^{12}$C/$^{13}$C = 70. We furthermore add a central compact component, in accordance to \citet{jorgensen2005a} who found a need for an abundance increase near the center. We describe this component with a jump in the abundance within the radius above 90 K ($\sim$100 AU) to $X(\textrm{HCN})=7\times 10^{-8}$.

The new addition to the model is a parameterization of the velocity field, which allows for both infall and rotation. This breaks the spherical symmetry of the model, making it two-dimensional, although we keep the spherical description of the density and temperature. We follow the approach presented by~\citet{brinch2008a} with a velocity field parameterized by two parameters,
\begin{eqnarray}
\mathbf{v}=\left (
\begin{array}{c}
	v_r \\ 
	v_\phi
\end{array}
\right ) = \sqrt{\frac{GM_*}{r}}
\left (
\begin{array}{c}
	-\sqrt{2} \sin{\alpha} \\ 
	\cos{\alpha}
\end{array}
\right ),
\end{eqnarray}
with $M_*$, the mass of the central object, and the angle $\alpha$ as free parameters. The velocity parameterization is such that $\alpha=0$ corresponds to Keplerian rotation and no infall, whereas $\alpha=\pi/2$ corresponds to free fall toward the center and no rotation. The use of this two-dimensional velocity field further introduces two free parameters, namely the inclination of the rotation axis with respect to the plane of the sky and the position angle of the rotation axis in the plane of the sky. We add a mean turbulent velocity field through a Doppler $b$-parameter of 0.2 km~s$^{-1}$ to the model. This value has been chosen to be large enough to get smooth line profiles, but small enough not to broaden the lines significantly. While this parameter is somewhat degenerate with the central mass in the sense that it affects the line widths, it cannot account for the different line widths in the interferometric and single-dish data. \textbf{The FWHM of the H$^{13}$CN lines are 2.1 km~s$^{-1}$ in the single-dish beam and 4.7 km~s$^{-1}$ for the lines meassured by the interferometer.} Only by letting the turbulent field vary across the model could we reproduce this effect. We have however chosen to use a constant turbulent field to keep the model simple, and a small change in its value does not affect our result significantly. For an in-depth discussion of turbulent line broadening we refer to, e.g., \cite{ward-thompson2001}. 

We use the molecular excitation and radiation transfer code \emph{RATRAN} \citep{hogerheijde2000a} to determine the level populations of the HCN molecule on a computational domain consisting of multiple nested regular grids. The grid resolves scales from 6 AU to $1.2\times 10^4$ AU, which is about the highest dynamical range of scales we can resolve with \emph{RATRAN} without loosing the gain of the accelerated lambda iteration (ALI) algorithm which makes the code feasible to run on a standard desktop computer. The resulting image cubes are afterward either convolved with the appropriate beam profiles for comparison to the single-dish lines or Fourier transformed for direct comparison to the visibility tables in the BIMA and SMA data sets.

\subsection{Model fit}
\begin{figure}
  \begin{center}
    \includegraphics[width=8.5cm]{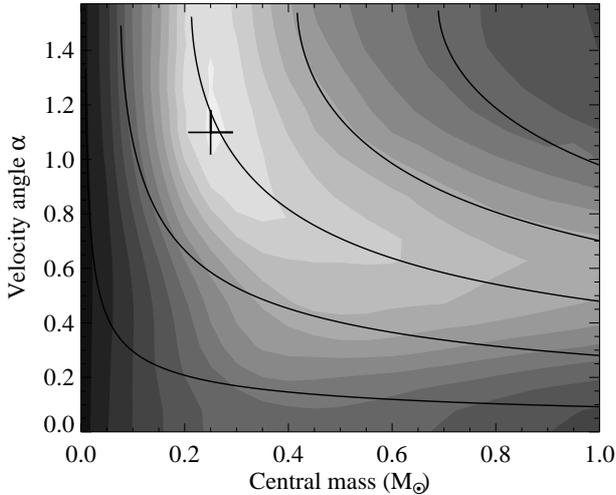}
  \end{center}
  \caption{H$^{13}$CN $\chi^2$ surface in the parameters $M_*$ and $\alpha$. The cross marks the minimum which lies at $M_*$=0.25 M$_\odot$ and $\alpha$=1.10 with a $\chi^2$ value of 1.84. The black lines are lines of constant accretion rate, with lines corresponding to 1, 3, 5, 7, and 9 $\times 10^{-5}$ M$_\odot$ yr$^{-1}$ going from the lowest line and up.}\label{iras2:chisquare}
\end{figure}

We run a grid of models in the parameter ranges $M_*\in \{ 0.0,1.0 \} $ and $\alpha \in \{ 0,\pi/2 \} $ and we compare the single-dish spectra and PV-diagrams from the SMA data of H$^{13}$CN only. A $\chi^2$-surface plot is shown in Fig.~\ref{iras2:chisquare} for an inclination of 40$^\circ$ {\bf and a position angle of 100$^\circ$}. We add the $\chi^2$ value of the two single-dish transitions, calculated channel by channel on a velocity range from -4 to +4 km~s$^{-1}$, to the $\chi^2$ value of the PV-diagram, where we compare pixels with a signal-to-noise better than three only. The $\chi^2$ is seen to be minimized for parameter values of $M_*=0.25$ M$_\odot$ and $\alpha=1.10$, which we take to be our best fit model. We have chosen to adopt an inclination parameter value of 40$^\circ$, because this parameter is not well-constrained by our fit. For any value below 40$^\circ$, we get an almost constant $\chi^2$ value and only for values higher than 40$^\circ$ we see a systematic increase in the $\chi^2$. The reason is that our best fit already favors an infall dominated model (i.e. almost spherical velocity field) and therefore the velocity field is mostly independent of the inclination. As we go toward models with more rotation, the fit rapidly deteriorates with increasing inclination, which only supports the result that little rotation is present in IRAS2A. Similarly, the position angle is unconstrained for the same reasons, in full agreement with the PV-diagrams shown in Sect.~\ref{iras2:results}. We have also plotted lines of constant mass accretion rate in Fig.~\ref{iras2:chisquare}. The accretion rate is calculated as
\begin{eqnarray}
\dot{M} = 4\pi r_0^2 \rho_0 \sqrt{\frac{2GM_*}{r_0}} \sin(\alpha)
\end{eqnarray}
where $\rho_0$ is the mass density at the distance $r_0=23\textrm{\ AU}$ The radius at which we evaluate the mass accretion rate has been chosen to be the same radius as was used by \cite{jorgensen2004b} for direct comparison. Our fit is seen to be consistent with a mass accretion rate of $5\times 10^{-5}$ M$_\odot$ yr$^{-1}$.

\begin{figure}
  \begin{center}
    \includegraphics[width=8.5cm]{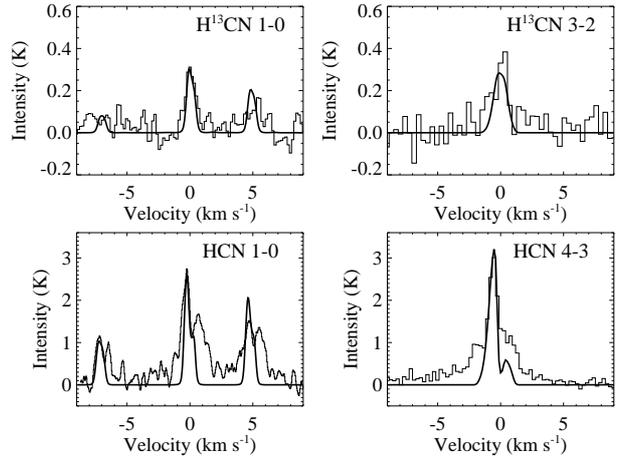}
  \end{center}
  \caption{Best model fit to the four single-dish transitions of HCN and H$^{13}$CN. }\label{iras2:singledish}
\end{figure}

The first fit, shown in Fig.~\ref{iras2:singledish}, is that of the single-dish lines H$^{13}$CN $J=$1--0, 3--2 and HCN $J=$1--0, 4--3. Being optically thin, the H$^{13}$CN lines are quite sensitive to the abundance profile and they constrain the radial variations well, as discussed by \citet{jorgensen2005a}. The $J=$1--0 lines show distinct hyperfine components. We model these components under the assumption of LTE between the hyperfine levels. The HCN lines have a much more distinct profile and they are much more sensitive to the velocity parameters. However, the large optical thickness of HCN and the possible contamination of outflow emission in these lines make it difficult to assess the quality of our model (not the fit) on basis of these lines. This is the reason why we do not include the HCN lines in our $\chi^2$ analysis. That said, our overall best fit model is also the model which gives the best fit to the HCN single-dish lines alone and it reproduces the ($u,v$)-amplitudes {\bf averaged over $\pm$~3~km~s$^{-1}$\ from the systemic velocity} nicely. Only the primary component of the H$^{13}$CN $J$=1--0 line is considered when calculating the $\chi^2$, but our best fit model produces a reasonable fit to the hyperfine components.

\begin{figure}
  \begin{center}
    \includegraphics[width=8.5cm]{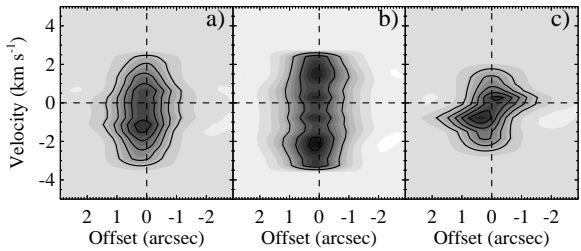}
  \end{center}
  \caption{Model PV-diagrams of H$^{13}$CN $J=$ 4--3 at
    P.A=10$^\circ$. \textbf{a)} Best fit model. \textbf{b)} High
    central mass model with $M_*$=1.0 M$_\odot$ and
    $\alpha$=1.1. \textbf{c)} Strongly rotating model with $M_*$=0.25
    and $\alpha$=0.2. All three panels are directly comparable to the
    center top panel of Fig.~\ref{pvs}. {\bf The contours levels are
      the same in all three panels and identical to those in
      Fig.~\ref{pvs}.}}\label{iras2:smafit}
\end{figure}

\begin{figure}
  \begin{center}
    \includegraphics[width=8.5cm]{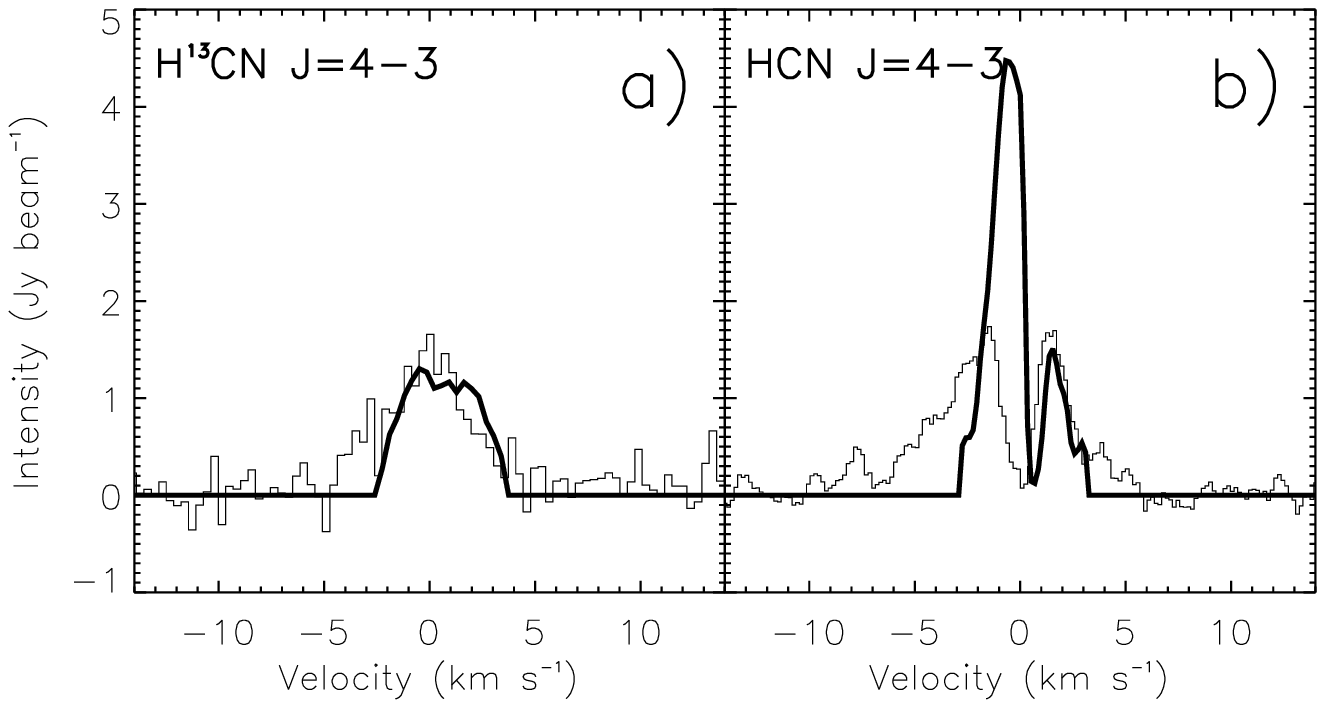}
    \includegraphics[width=8.5cm]{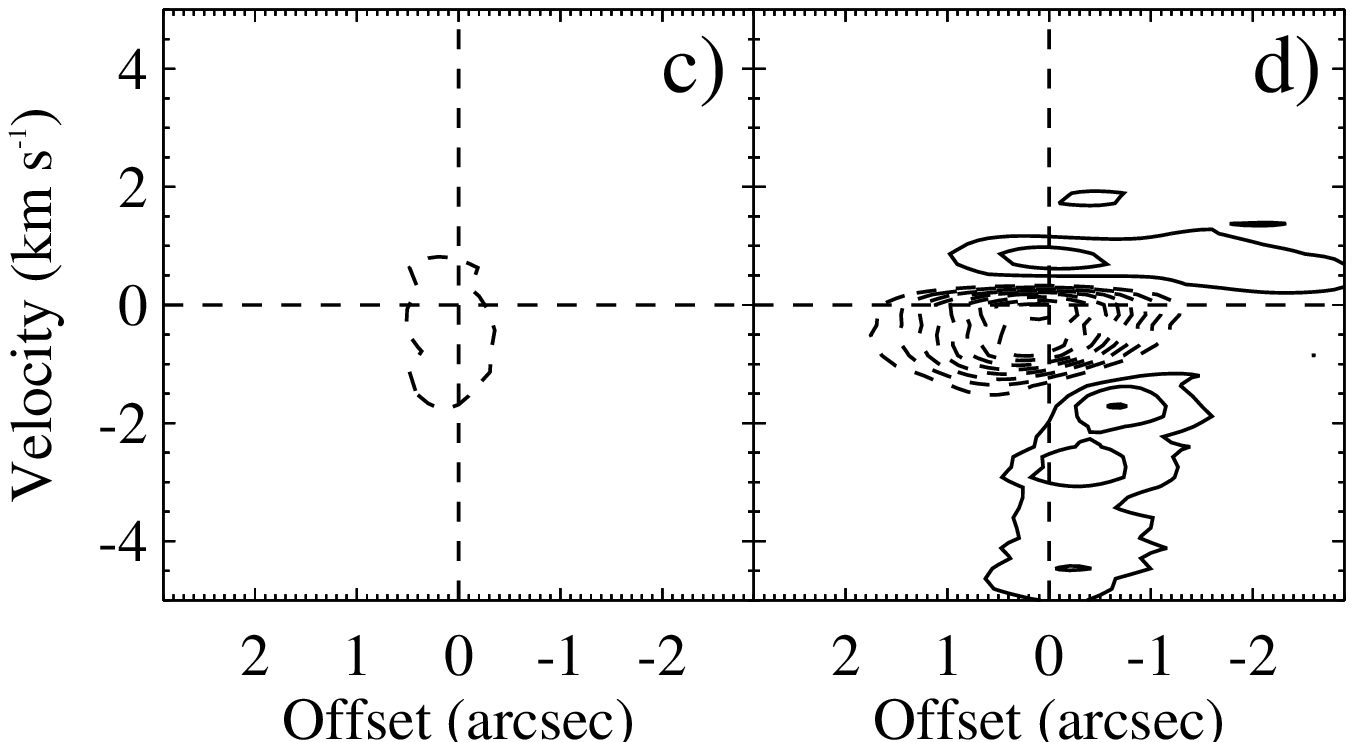}
  \end{center}
  \caption{Fit to the SMA data. Panel \textbf{a)} and \textbf{b)} show a single spectrum toward the center while panel \textbf{c)} and \textbf{d)} show the residual of the model fit to the PV-diagrams.}\label{iras2:smahcn}
\end{figure}

Figure \ref{iras2:smafit} shows the best fit H$^{13}$CN model PV-diagram (panel a) as well as two other models for comparison, $M_*$=1.0 M$_\odot$ and $\alpha$=1.1 (panel b) and $M_*$=0.25 M$_\odot$ and $\alpha$=0.2 (panel c). The contour levels in these plots are set to the same values as the levels in Fig.~\ref{pvs} and so these plots are directly comparable to the center panel (PA=100$^\circ$) of that figure. The best fit model is seen to reproduce the features of the data well in terms of line intensities and emission distribution, while the model in panel b has too weak lines and the model in panel c shows a skewed emission distribution. The asymmetry seen in panel c is typical for a rotating velocity field and this feature is not seen at all in the data. A direct comparison between the SMA data and our model is shown in Fig.~\ref{iras2:smahcn}. Panels a) and b) show the fit to a single spectrum toward the center position of IRAS2A for H$^{13}$CN and HCN respectively. The fit to the H$^{13}$CN spectrum reproduces the entire line profile well. From the residual of the fit to the PV-diagrams (panel c) it can be seen that it is not only the center position that fits well. In the case of HCN, however, the best fit model does not work as well. While the red side of the spectrum is relatively well reproduced, also spatially, the blue side suffers from a too strong peak and a too narrow line wing. The line wing emission is most likely the outflow, similar to the wing seen in HCN $J$=4--3 in Fig.~\ref{iras2:singledish}. The strong asymmetry, predicted by our model, is not seen in the SMA spectrum, where the blue and the red sides peak at the same intensity. Because we do see the asymmetry in the single-dish lines, the missing blue peak is likely due the HCN line becoming optically thick on large scales where the interferometer filters out emission {\bf --- and where our model does not necessarily apply, e.g., due to departures from the simple spherical power-law density profile or associated temperature profile.}

\subsection{Outflow}
\begin{figure}
  \begin{center}
    \includegraphics[width=8.5cm]{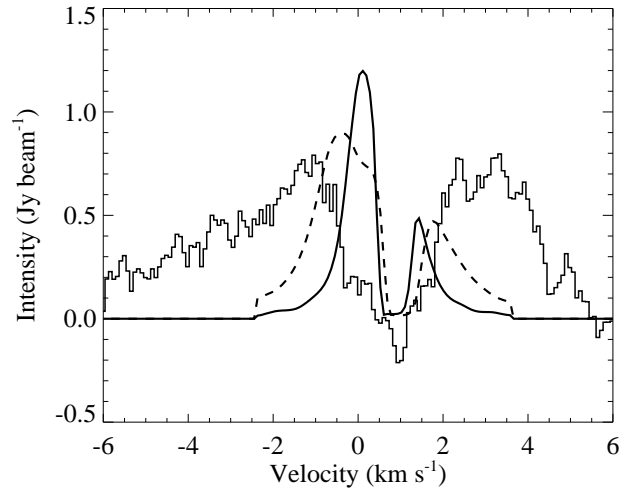}
  \end{center}
  \caption{A HCN $J=$ 1--0 spectrum toward the center of IRAS2A as observed by BIMA. Two models are overplotted; our best fit model in solid line style and a model with $M_*$=1.0 M$_\odot$ in dashed line style. The truncation of the model line wings at 2.5 km~s$^{-1}$ is due to the finite pixel size of our grid. }\label{iras2:bimaspec}
\end{figure}

IRAS2A is known to drive two strong outflows, along axes that are almost perpendicular to each other. We do not model these outflows, and therefore we do not expect to reproduce features in the data that are associated with the outflows. As an optically thin line, we expect that H$^{13}$CN is much less affected than HCN, and indeed only the single-dish HCN lines show broad wings with excess emission not accounted for by our model. Both the single-dish and the SMA HCN spectra are contaminated by outflow emission, and if we go to the BIMA observations of the low excitation line $J=$ 1--0, it immediately becomes clear that care needs to be taken when modeling the HCN lines. Figure \ref{iras2:bimaspec} shows the BIMA spectrum toward the center of the source. Two models are overplotted, the best fit with a solid line and a model with $M_*$=1.0 M$_\odot$ using a dashed line style. The observed spectrum is extremely broad with wings extending all the way to $\pm$6 km~s$^{-1}$. Our model does not reproduce these velocities, and indeed, even if we increase the mass significantly, the line width only increases a little.  

\section{Discussion}\label{iras2:disc}
{\bf Within the framework developed on basis of the hydrodynamical simulations \citep{brinch2008a},} we determine the central mass of IRAS2A to be 0.25 M$_\odot$ with a $1\sigma$ uncertainty of $\pm$0.1 M$_\odot$. This value is considerably smaller than the mass of the envelope of 1.7 M$_\odot$ {\bf obtained from modeling of the dust continuum observations of IRAS2A and} in this sense, IRAS2A is a true Class 0 object with $M_*<$ $M_{\rm{env}}$. It should be noted that the implicit assumption for the velocity field in this work is that material is free falling --- if it is not rotating. It is possible that the circumstellar material is not in free fall, either because free fall velocity has not yet been reached or because material is being supported magnetically or otherwise. If this is the case, the mass parameter underestimates the true mass of the star by up to a factor of four, depending on how far the system is from free fall~\citep{brinch2008a}. 

Figure \ref{iras2:chisquare} shows that our best fit is consistent with a mass accretion rate of 5$\times$10$^{-5}$ M$_\odot$ yr$^{-1}$. If the mass accretion rate has been constant, the current central mass would have been accreted in only 5$\times$10$^3$ yr. However, the accretion rate may have been lower in the past or maybe most of the material does not accrete directly onto the star. The age of the star, which we derive from kinematical arguments, is consistent with previously estimated time scales for IRAS2A. Time scales based on the outflow dynamics have been estimated by \cite{jorgensen2004} and \cite{bachiller1998} to be 4$\times$10$^3$ and (3--7)$\times$10$^3$ yr, respectively. \cite{jorgensen2004b} found a time scale of 1.7$\times$10$^4$ yr from an inside-out collapse model fitted to single-dish observations{\bf, with an accretion rate about an order of magnitude lower than what we find here. In contrast to the model of~\cite{jorgensen2004}, which is derived from observations probing the larger-scale non-collapsing envelope, our high angular-resolution data probe the core kinematics on scales of a few hundred AU. The difference between these results therefore also suggests the limit to the usability of the simple self-similar collapse models for these earliest stages of low-mass protostars.}




{\bf If the bolometric luminosity of IRAS2A of 20--25 L$_\odot$ originates solely} from the accretion through the relation $L=GM_*\dot{M}/r$, the material needs to accrete at a radius of approximately 16--20 solar radii (0.07--0.09 AU), which is about a factor of 6--8 bigger than the estimated stellar radius of young protostars \citep{stahler1988} and within the anticipated radius of a disk. {\bf Still, the earlier interferometric studies of the continuum emission of IRAS2A show that the larger-scale infalling envelope cannot explain the continuum emission on scales of a few $\times$ 100~AU, but that an additional compact component has to be present there \citep{looney2003,jorgensen2004,jorgensen2005}. If this compact component indeed reflects the presence of a circumstellar $\sim 0.1$~$M_\odot$ disk, it must} also have been built up by the accretion flow within the lifetime of the object. It therefore appears plausible that matter is not accreted directly onto the central star.

{\bf The question is why we do not see the rotation if the compact continuum component is interpreted as a disk. One possible explanation of the absence of a rotational velocity component is that the disk lies exactly in the plane of the sky and we are viewing it face-on. This however, does not agree with the morphology of the two outflows which both have components in the plane of the sky. Another possibility is that it is not a stable disk yet but rather a pile up of low-angular momentum material very close to the star, which is still dominated by infall. Such a structure could represent a ``pseudo-disk'' found in the models of collapse of magnetized cores by \cite{galli1993}. By contraction, this material could still settle in a rotating disk on a ``local dynamical time-scale'' - essentially the free-fall time-scale. At a radius corresponding to the SMA beam (0.5$''$ = 110 AU) this time-scale is a few $\times 10^3$~yr, comparable to the time-scale found above. Such low angular momentum material may therefore not have settled in a rotationally supported disk, yet. Finally, since IRAS2A drives two outflows, it is reasonable to assume that IRAS2A is a close binary. If this is the case, the compact dust component may lie inside the tidal truncation radius, preventing material from settling into a Keplerian, rotationally supported disk \citep{artymowicz1994}.}

{\bf Still, the main conclusion remains that IRAS2A appears to be} an example of a very young protostar based on our best fit velocity model, with both the velocity field parameters pointing toward a young source. Its velocity field is characterized by a high degree of infall relative to rotation, which in comparison to the study of the velocity field in hydrodynamical simulations of star formation by~\cite{brinch2008a} suggests a young source. In that paper it was shown that the $\alpha$ parameter evolves smoothly with time from a high value ($\pi$/2) to a low value (0.0) as a core collapses and a disk is formed. A study similar to this one was done by \cite{brinch2007a} for another young star, L1489~IRS, classified as a Class I source. For this source, a disk was detected based on similar kinematical arguments as we use here, and for that source an $\alpha$ value of 0.26 was found. Based on the value of the $\alpha$ parameter alone, we conclude that IRAS2A is a younger source than L1489~IRS, and this is entirely consistent with the general perception that Class 0 objects are younger than Class I objects. 

\section{Summary and outlook}\label{iras2:outlook}
In this paper we have modeled the velocity field of the Class 0 young stellar object NGC1333--IRAS2A. We have constrained our model
primarily by the PV-diagram of the H$^{13}$CN emission observed at high angular resolution with the SMA. Despite the angular resolution
we find no evidence for Keplerian motions on any scales, even though a very dense, compact component, interpreted as a disk, is needed on
scales of a few$\times$100 AU to reproduce the ($u,v$)-amplitudes, seen in the continuum data.

Repeating this approach for a sample of protostars would allow statistical comparison of the velocity fields of these stars and maybe allow a new evolutionary classification based on $\alpha$ and $M_*$/$M_{\rm{env}}$. This will only be possible when the Atacama Large Millimeter Array (ALMA) comes online, because ALMA will be able to do observations like the ones presented here in snapshot mode as opposed to a full night per source needed by the SMA. Furthermore, ALMA will allow us to zoom in on the innermost parts of protostellar objects and tell us what happens to the velocity field as circumstellar disks form. In this paper we did not model the outflow emission that is clearly seen in the BIMA HCN $J$=1--0 line and also noticeable in the HCN single-dish lines. By using the SMA to observe high-excitation lines, we filter out a lot of the emission on large scale where we expect the outflow to be dominating the velocity field. Still, we do not fit the HCN $J$=4--3 line very well. This illustrates the need of good radiation transfer outflow models which can be used to investigate the interaction between accretion and outflows on very small scales.\\

\noindent \emph{Acknowledgments} The authors would like to thank Ewine van Dishoeck and Phil Myers for useful comments and discussion about the manuscript. CB is partially supported by the European Commission through the FP6 - Marie Curie Early Stage Researcher Training programme. The research of MRH is supported through a VIDI grant from the Netherlands Organization for Scientific Research.\\

\bibliographystyle{aa}
\bibliography{references}
\end{document}